# Capacitance characterization of Graphene/n-Si Schottky junction solar cell with MOS capacitor

Masahiro Teraoka, Yuzuki Ono, Hojun Im[*]

*Graduate School of Science and Technology, Hirosaki University, Hirosaki 036-8561, Japan*

**ABSTRACT**

  We have demonstrated a simple and accurate method for characterizing the capacitance of Graphene/n-Si Schottky junction solar cells (GSSCs) which embed the metal-oxide-semiconductor (MOS) capacitor. We measured two types of GSSCs, one with thermal annealing treatments (w-a) and one without (wo-a). It was found that the wo-a GSSC exhibits a two-step feature in the phase versus forward bias voltage relationship, which may be attributed to the presence of polymethyl methacrylate residues. By considering the capacitance of the MOS capacitor ($C_{mos}$) and its standard deviation, we successfully obtained the capacitance of the Schottky junction ($C_{Sch}$), and evaluated meaningful built-in potentials (Schottky barrier heights) which are 0.51V (0.78eV) and 0.47V (0.75eV) for the w-a and wo-a GSSCs, respectively, by the Mott-Schottky analysis. We also briefly discuss the relationship between $C_{Sch}$ and the Nyquist and Bode plots, finding that the RC time constant decreases due to the subtraction of $C_{mos}$.

# 1. Introduction

Graphene/n-Si Schottky junction solar cells (GSSC) have attracted attention as a promising solar cell technology because of their low-cost fabrication and high-power conversion efficiency (PCE).[1]–[8] The PCE of GSSCs has increased rapidly from approximately 1.5% in 2010 to around 15% in 2015, using various strategies such as hole doping, silicon surface passivation, and anti-reflection coating.[9]–[18] However, the PCE has plateaued in recent years, with only slight improvements observed. The latest reported values show a PCE of around 17%.[19] To achieve a breakthrough, it is essential to understand the underlying mechanisms of GSSCs, requiring well-established methods which evaluate and analyze the performance via photovoltaic parameters: PCE, short current density ($J_{SC}$), open circuit voltage ($V_{OC}$), the recombination lifetime, charge extraction time, etc. Capacitance characterizations have provided abundant information on the steady-state properties in solar cells.[20],[21] Particularly, capacitance vs. bias-voltage ($C$-$V$) characteristics have been facilitated to estimate the Schottky barrier height (SBH) and the dop concentration ($N_d$) using the so-called Mott-Schottky plot.[10],[22]–[24] Moreover, capacitance spectra have been used to evaluate the interfacial properties via Nyquist plot (and/or Bode plot) combined with the equivalent circuit model such as the Randle circuit.[20] However, the relevant analysis methods have not yet been standardized for GSSCs, which cause the difficulties in comparative analysis between research groups.

To resolve the issues, there have been several investigations of the capacitance characteristics, focusing on the graphene/n-Si Schottky junction diodes. For instance, C. Yim et al. separately evaluated the capacitive components such as the Schottky junction, metal electrodes, and extract the more reliable SBH value in impedance spectroscopy, using a proper equivalent circuit model.[25] S. Riazimehr et al. have extract more accurate photovoltaic parameters such as the built-in potential from the $C$-$V$ characteristics, considering the metal-oxide-semiconductor (MOS) capacitor.[26] I. Matacena et al., have elucidated the forward bias capacitance to monitor graphene/silicon interfaces.[27] These considerations should be also appropriately applied to the evaluation of GSSCs.

Recently, we have reported that the polymethyl methacrylate (PMMA) residues deteriorate the efficiency of GSSCs due to the charge trap and the increase of the recombination rate. And the PMMA residues can be removed by the thermal annealing treatments in addition to the acetone solution, enhancing the performance of GSSCs.[28] Here, we demonstrate a facile and more accurate analysis methods of capacitance characteristics in GSSCs with (w-a) and without the thermal annealing (wo-a) treatment, considering the MOS capacitor embedded in GSSCs.

## 2. Experimental details

To fabricate GSSCs, we first grew graphene on catalytic Cu foil using low-pressure chemical vapor deposition. We then transferred the graphene onto a patterned n-Si substrate (approximately 10×10 mm$^2$ in size, with a thickness of 200 μm and a resistivity of 1-10 Ω) using a conventional PMMA-assisted method, following the process described in our previous paper.[28] Figure 1(a) shows the completed solar cell. To accurately evaluate the photovoltaic properties of the GSSC, we also fabricated MOS capacitors using a similar process as the GSSC, but without the active window as shown in Fig. S1.

Capacitance measurements have been carried out on both the w-a and wo-a GSSCs, using an LCR meter (ZM2376, NF Corp.) automated by the LabVIEW program. To obtain *C-V* characteristics, we measured capacitance as a function of bias-voltage in the range of -4.5 to 1 V at 10 kHz. To obtain capacitance spectra, we measured capacitance in the range of frequency from 1 to 1M Hz at zero bias. Capacitances were measured in the parallel mode of the LCR meter, with an AC signal level perturbation fixed at 20 mV throughout all measurements. Additionally, we measured current density vs. voltage (*J-V*) characteristics in both the dark and under illumination using a source meter unit (Keithley2400, Tektronix) and a xenon lamp with an AM1.5G filter as a 1-sun light source (100 mWcm$^{-2}$).

## 3. Results and discussion

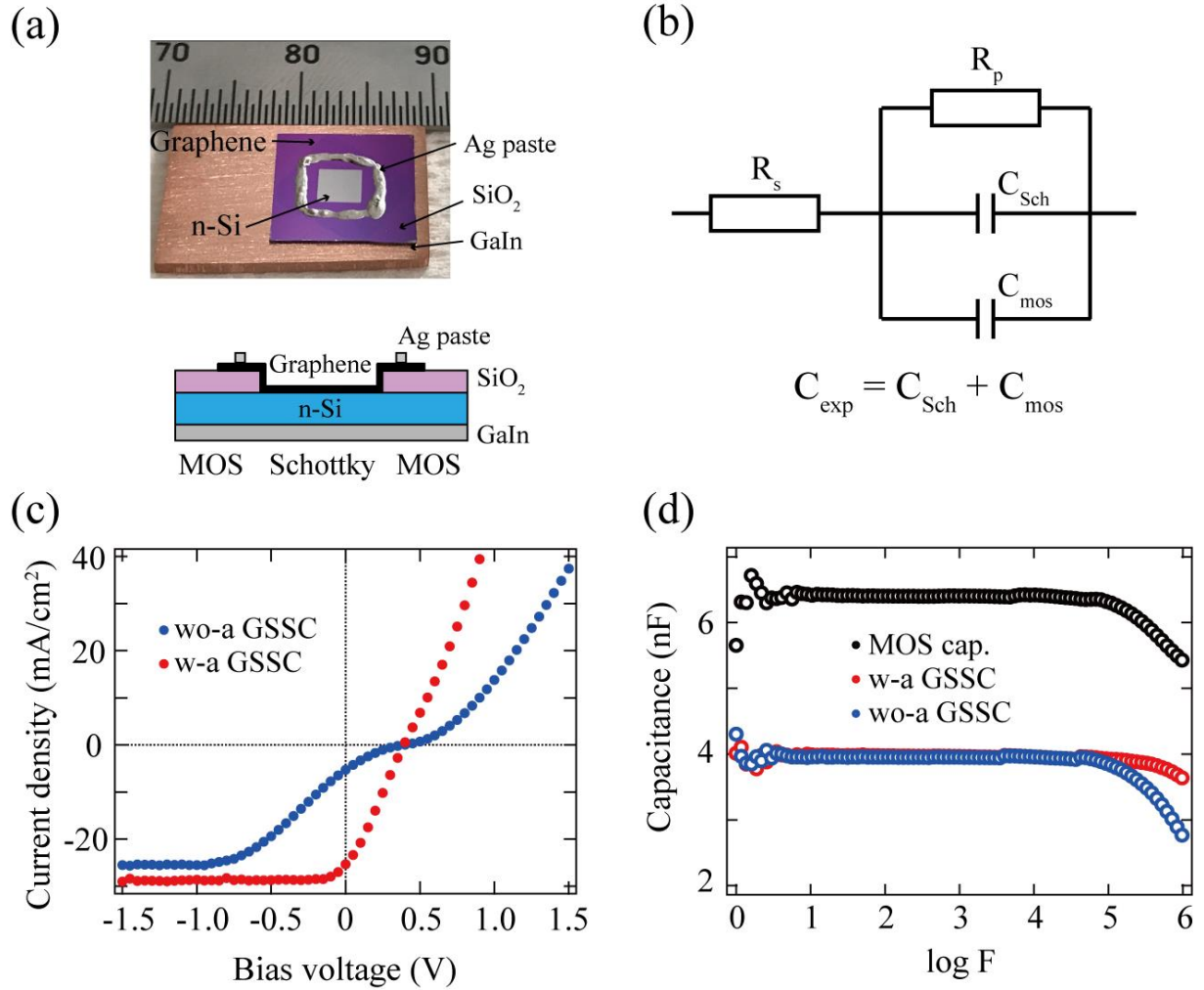

**Fig. 1.** (a) Photograph of the fabricated GSSC (upper part) and its schematic in the side view (lower part). (b) Corresponding equivalent circuit model. (c) *J-V* characteristics of w-a and wo-a GSSCs in illumination. (d) Capacitance spectra of the GSSCs and the MOS capacitor (Fig. S1) from 1 to 1 MHz. Capacitance of the MOS capacitor was normalized to the area of 1 cm$^2$ (nF/cm$^2$).

In capacitance analysis, the solar cell is often interpreted using the so-called Randle circuit, which includes a parallel capacitor ($C_p$), parallel resistor ($R_p$), and series resistor ($R_s$). However, in actual devices, GSSCs include an additional MOS capacitor (Gr/SiO$_2$/n-Si) to electrically separate the n-Si and graphene for the anode electrode, as shown in Fig. 1(a). Parts of the MOS capacitor can be considered parasitic components in parallel with the Schottky junction (Gr/n-Si) as depicted in the equivalent circuit of Fig. 1(b). These GSSC geometries must be considered to accurately evaluate device performance, as previously mentioned in the introduction.[26] To develop such accurate capacitance characterizations, demonstrations with well-established GSSCs are essential.

Figure 1(c) shows the *J-V* characteristics of the w-a and wo-a GSSCs. In our previous work, it has been reported that PMMA residues in GSSCs can deteriorate efficiency due to charge trapping and increase recombination rates. The PMMA residues can be effectively removed by thermal annealing treatments in addition to acetone solution, enhancing GSSC performance.[28] This work successfully reproduced these phenomena as shown in Fig. 1(c) and Fig. S2. Photovoltaic parameters such as short-circuit current density ($J_{sc}$), open-circuit voltage ($V_{oc}$), filling factor (FF), power conversion efficiency (PCE), and ideality factor (n) were estimated to be 25.4 (5.6) mA/cm$^2$, 0.39 (0.40) V, 0.28 (0.18), 2.8 (0.4) %, and 1.16 (1.35) for the w-a (wo-a) GSSCs, respectively. Such performances are comparable to those of the previous reports as the pristine devices.[9,28] Furthermore, it has been well-known that the capacitance of GSSCs can be successfully analyzed by the Mott-Schottky plot. These conditions allow us to effectively demonstrate the capacitance characteristics by comparative studies.

To investigate the influence of the MOS capacitor, we separately fabricated a MOS capacitor of Gr/SiO$_2$(500 nm)/n-Si(200 μm) with the same geometry to the used GSSCs, except for the absence of the active area in center as shown in Fig. S1. Fig. 1(d) displays the capacitance spectra of the w-a (wo-a) GSSCs and the MOS capacitor in the frequency range of 1 Hz to 1 MHz. It was found that all capacitance spectra exhibited similar behaviors, displaying a plateau in 10 – 100 kHz, indicating the good formation of the depletion region. This is an important condition to study the capacitance characteristics. In the higher frequency range of 100 kHz to 1 MHz, we observed several patterns in the reduction of capacitance. This behavior may be attributed to the capacitance caused by electrode contact, which is highly sensitive to each device. In the low frequency range below 10 Hz, while the signal-to-noise ratio slightly increases due to the limitations of the equipment for low-current detection, we can still observe a tendency in the capacitance spectra. Therefore, in this study, we evaluate the *C-V* characteristics at a frequency of 10 kHz and consider the capacitance spectra obtained between 1 and 100 kHz.

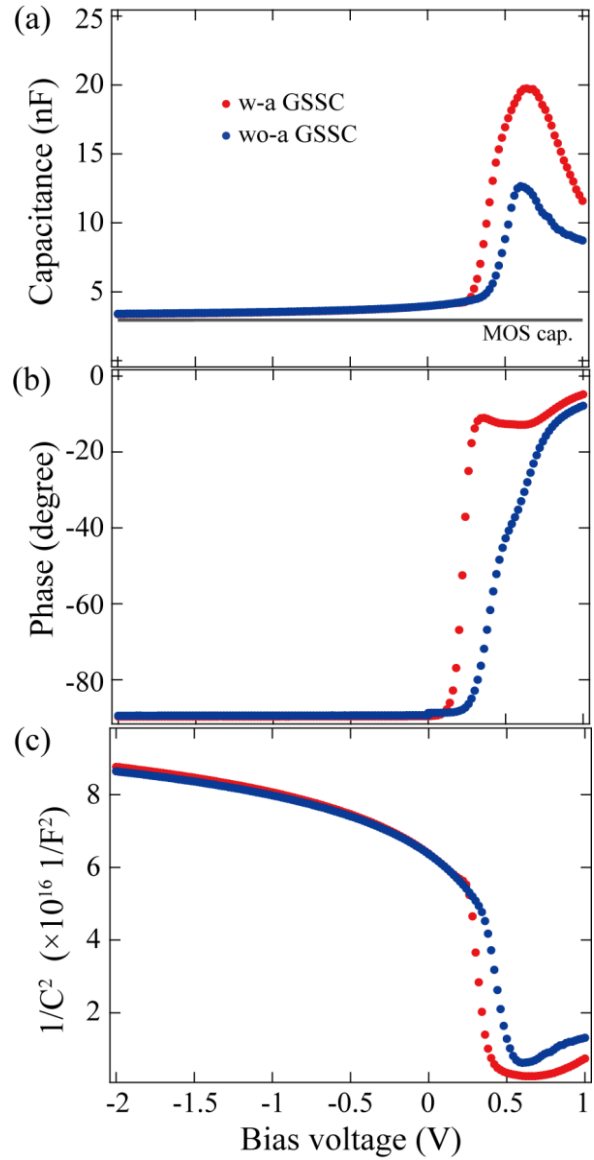

**Fig. 2.** (a) *C-V* characteristics of the w-a and wo-a GSSCs obtained at 10 kHz. Capacitance of the MOS capacitor normalized to the actual area of the MOS capacitor in the solar cells. (b) Phase vs. the bias voltage plot. (c) Mott-Schottky plots obtained from $C_{exp}$ of (a).

Figure 2(a) shows the *C-V* characteristics of the w-a and wo-a GSSCs in both the reverse and forward bias voltage (from -2 to 1 V) at 10 kHz. Capacitance of the MOS capacitors ($C_{mos}$), which are normalized to the actual area of the MOS capacitor embedded in GSSCs, are also displayed (gray lines). $C_{mos}$ shows a monotonic behavior and little influences on qualitative interpretation of the capacitance behaviors of GSSCs. In the Schottky junction, capacitance is mainly attributed to space charges of the depletion layer in the reverse bias region, while that to thermionic emission carriers in the forward bias region. On the other hand, the MOS capacitor generally shows different behaviors in the accumulation (forward bias), depletion (near zero bias), and inversion (reverse bias) regions. To accurately evaluate capacitances, we need to simultaneously consider all of these contributions in both the Schottky junction and MOS capacitor. Capacitances of both the w-a and wo-a GSSCs show similar behaviors in -2 to ~0.2 V across zero bias voltage, slowly increasing from ~3.4 to ~4.2 nF. Moreover, their phases are almost -90° as shown in Fig. 2(b), indicating that the capacitor properties are dominant. This implies that the slow increase of capacitance comes from the decrease of the space charge width. On the other hand, capacitances rapidly increase in the above 0.3 V and decrease in the above ~0.7 V, showing the peak structure. The large increase of capacitance originated from the exponential increase of the thermionic current. Above 0.7 V, capacitances decrease due to the increasing influence of $R_s$ as the forward bias voltage increases.[20),27),29)] This shows good agreement with the results of the J-V characteristics (Fig. 1(a) and Fig. S1(b)). It is worth noting that the w-a and wo-a GSSCs exhibit different peak structures; the rising edge is lower voltage in the w-a GSSC than in the wo-a GSSC, and the maximum capacitance value is larger in the w-a GSSC, showing broad feature. Particularly, the phase vs. bias voltage plots in Fig. 2(b) reveal clear differences between the two types of GSSCs. In the wo-a GSSC, the phase exhibits a two-step feature, with a stair around -40 degrees at ~0.4 V followed by a rise to 0 degrees at ~0.8 V. In contrast, in the w-a GSSC, the phase suddenly changes from -90 to 0 degrees at ~0.3 V. These findings suggest that the w-a GSSC has a larger thermionic current than the wo-a GSSC, resulting in a better ideality factor that is consistent with the *J-V* characteristics. The two-step phase feature observed in the wo-a GSSC can be attributed to capacitive behavior, possibly due to trapped charge carriers in the PMMA residues.

In addition to the above qualitative understanding, *C-V* characteristics also allow us to quantitatively evaluate the SBH and $N_d$ via the Mott-Schottky plot, which is $1/C^2$ vs $V$ as described in Eq. (1) and (2).[30)]

$$\frac{1}{C^2} = \frac{2}{q\varepsilon\varepsilon_0 N_d}(V_{bi} - V) \qquad (1)$$

$$SBH = V_{bi} + \frac{kT}{q} \ln\left(\frac{N_c}{N_d}\right) \qquad (2)$$

where *q* is the elementary charge, *ε* the relative permittivity (11.7 for Si), $\varepsilon_0$ the permittivity of free space, $V_{bi}$ the built-in potential, and $N_c$ the effective density of conduction band ($2.8 \times 10^{19}$ cm$^{-3}$ for Si).[30)]

Figure 2(c) shows the Mott-Schottky plots obtained from $C_{exp}$, but these plots exhibit non-meaningful behavior. This is because $C_{exp}$ includes not only the Schottky junction ($C_{Sch}$) but also the MOS capacitor ($C_{mos}$) in parallel (Fig. 1(b)). To address this issue, we here propose a simple and accurate method as follows.

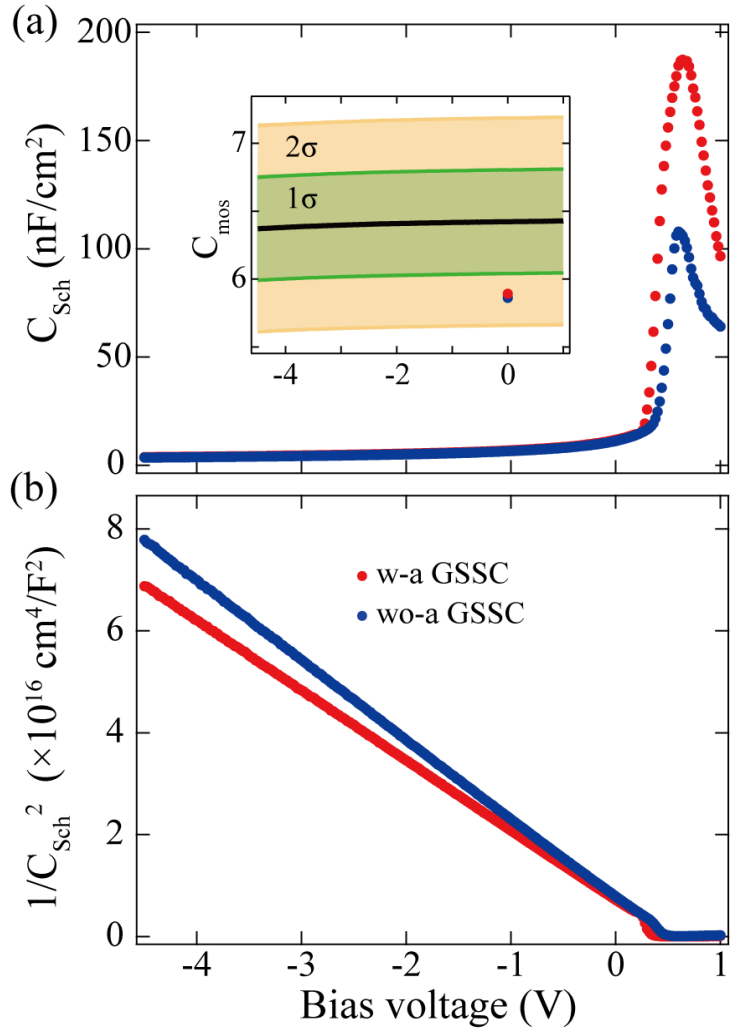

**Fig. 3.** (a) $C_{Sch}$-$V$ characteristics of the w-a and wo-a GSSCs obtained from Eq. (3). $C_{Sch}$ and $C_{mos}$ of the inset were normalized to the area of 1 cm$^2$. The values of ($C_{mos} + \Delta C$) at zero bias were marked by solid circles in the inset. (b) Mott-Schottky plots obtained from $C_{Sch}$ of (a).

The inset of Fig. 3(a) displays the average $C_{mos}$, obtained from 7 MOS capacitors, as a function of bias voltage, and normalized to an area of 1 cm$^2$. The standard deviations ($\sigma$) are also shown alongside the data. The MOS capacitor has the same geometry as the GSSC, except for the absence of the Schottky junction area (Fig. S1(a)). Capacitance monotonically increases with the bias voltage from ~6.37 to ~6.43 nF/cm$^2$, showing a shallow uphill slope. And the 1$\sigma$ value is estimated to be around 0.38 nF/cm$^2$. It should be noted that the capacitance variation caused by the transition from the inversion to the accumulation region via the depletion regions does not clearly appear due to the thick 500 nm SiO$_2$ layer. To extract the $C_{Sch}$ values, $C_{mos}$ should be subtracted from the $C_{exp}$ values. It should also be noted that each individual device may have slight differences due to the delicate and sensitive fabrication processes, making capacitance characterization analysis challenging. Therefore, we demonstrate quantitative analysis using the average $C_{mos}$ with standard deviation, as described in Eq. (3).

$$C_{Sch} = C_{exp} - (C_{mos} + \Delta C) \qquad (3)$$

where $\Delta C$ the adjustable capacitance based on the standard deviation of $C_{mos}$.

Fig. 3(a) shows the $C_{Sch}$-$V$ characteristics extracted from the w-a and wo-a GSSCs using Eq. (3). We plotted the Mott-Schottky plots obtained from $C_{Sch}$ in Fig. 3(b), using $\Delta C$s of -0.54 and -0.57 nF/cm$^2$ for the w-a and wo-a GSSCs, respectively. Note that the $\Delta C$ values are nearly 1$\sigma$ as marked in the inset of Fig. 3(a). The Mott-Schottky plots have revealed that 1/$C^2$ is excellently proportional to the bias voltage. We estimated the built-in potentials of the w-a and wo-a GSSCs to be ~0.51 and ~0.47 V, respectively, from the intercept of the $x$-axis (bias voltage). Additionally, using Eq. (2), we evaluated the SBHs to be 0.78 and 0.75 eV for the w-a and wo-a GSSCs, respectively, which is ~0.1 eV less than those of $J$-$V$ characteristics (0.85 eV and 0.92 eV for the w-a and wo-a GSSCs, respectively). This behavior has been previously observed in graphene-silicon diodes and attributed to a larger effective Richardson constant (n-Si for 112 Acm$^{-2}$K$^{-2}$) than the actual value. These indicate that our proposed analysis methods are working well despite the slightly different values between $C$-$V$ and $J$-$V$ characteristics. To confirm the validity of the above method, we additionally applied this method to 10 devices of GSSCs (shown in Fig. S3) and found that the average SBH value is 0.774 eV ± 0.034 eV, which is quite reasonable compared to the previous results. The distribution of $\Delta C$ is within 2$\sigma$, which corresponds to around 10% of the $C_{mos}$ values. We also estimated the $N_d$ of the w-a and wo-a GSSCs to be 8.8×10$^{14}$ and 7.7×10$^{14}$ cm$^{-3}$, respectively, from the slope of the Mott-Schottky plot in Fig. 3(b), which are proper values for the dop concentration of the used n-Si

(1–10 Ω).[30)]

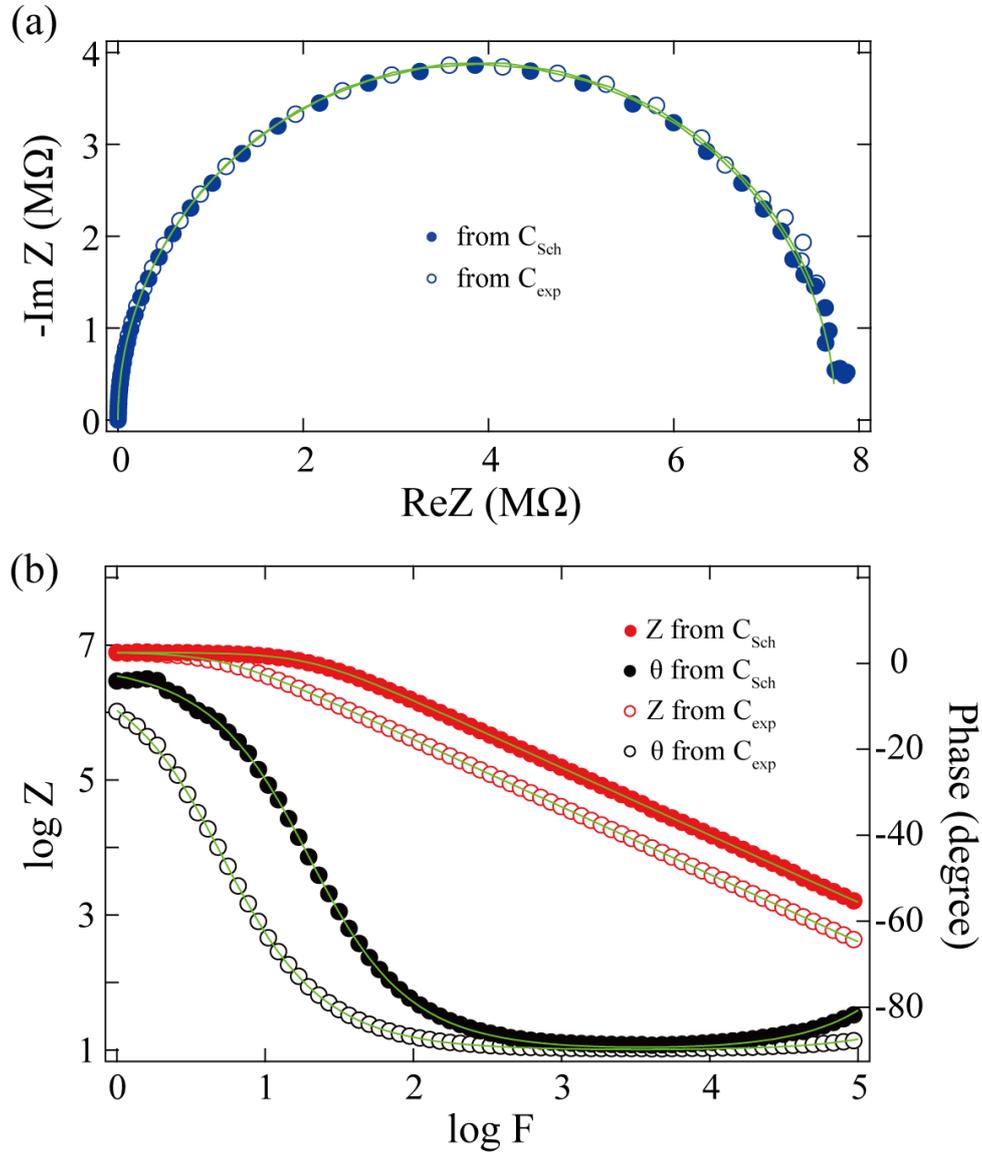

**Fig. 4.** (a) Nyquist and (b) Bode plots of the w-a and wo-a GSSCs obtained from the $C_{exp}$ and $C_{Sch}$. The $C_{Sch}$ values can be regarded as those for the active area (3×3 mm$^2$). The solid lines are fitted by the Randle circuit model.

Finally, we would like to briefly discuss the relationship between the proposed analysis and the Nyquist plot and Bode plot, using the w-a GSSC dataset. We can easily expect that the RC time constant would decrease due to the reduction of capacitance from $C_{exp}$ to $C_{Sch}$, assuming that $R_p$ does not significantly change. Under this condition, we can obtain the Nyquist and Bode plots through a simple data analysis, using Eqs. (4) and (5). We use $C_{Sch}$ instead of $C_{exp}$ to represent $C_p$.

$$|Z| = \frac{R_p}{\sqrt{1 + (\omega C_p R_p)^2}} \quad (4)$$

$$\vartheta = -arctan(\omega C_p R_p) \quad (5)$$

where $Z$ and $\theta$ are impedance and phase, respectively.[31]

We performed impedance spectroscopy measurements in the dark to study carrier transport under simplified conditions. Figures 4(a) and 4(b) show Nyquist and Bode plots, respectively, obtained from $C_{exp}$ and $C_{Sch}$. The Nyquist plots were fitted by the simple Randle circuit model using EIS spectrum analyser.[32] The values of $C_{exp}$ ($C_{Sch}$), $R_s$, $R_p$, and $C_p$ were estimated to be 17.7 (248) Ω, 7.79 (7.75) MΩ, and 3.96 (1.04) nF, respectively, revealing a decrease in the RC time constant from 31 to 8.5 msec. The reduction of the $RC$ time constant was also observed in the Bode plot. However, we refrain from delving into its underlying physical implications due to limited investigation at this stage. A more comprehensive understanding requires further study. For the wo-a GSSC, Nyquist and Bode plots mainly exhibit the capacitance behavior due to the large $R_p$ value, as shown in Fig. S4.

## 4. Summary

We propose a simple and accurate method for characterizing the capacitance of GSSCs. Our method considers $C_{mos}$ with an adjustable $\Delta C$ based on the experimental standard deviation. Capacitance measurements reveal that the Mott-Schottky plots from the $C_{exp}$ data show no meaningful features due to the inclusion of $C_{mos}$. On the other hand, the Mott-Schottky plots from the $C_{Sch}$ data exhibit reasonable features, allowing us to evaluate the built-in potentials (SBH) as 0.51 (0.78 eV) and 0.47 V (0.75 eV) for the w-a and wo-a GSSCs, respectively. The estimated $N_d$ values of $8.8 \times 10^{14}$ and $7.7 \times 10^{14}$ cm$^{-3}$ for the w-a and wo-a GSSCs, respectively, are also relevant for those of the 1–10 Ω n-Si substrates used. These results indicate the effectiveness and usefulness of our proposed method. We also briefly discuss the relationship between $C_{Sch}$ and the Nyquist and Bode plots. If $R_p$ were not changed, we could obtain the Nyquist and Bode plots using $C_{Sch}$ via simple data analysis. We found that the $RC$ time constant becomes small due to the subtraction of $C_{\text{mos}}$ from $C_{\text{exp}}$. We believe that our proposed method provides a comprehensive approach to understanding the mechanism in GSSCs.

REFERECE

1) F. Bonaccorso, L. Colombo, G. Yu, M. Stoller, V. Tozzini, A. C. Ferrari, R. S. Ruoff and V. Pellegrini, Science **347** [6217], 1246501 (2015).
2) A. Di Bartolomeo, Phys Rep **606**, 1 (2016).
3) M. F. Bhopal, D. W. Lee, A. ur Rehman and S. H. Lee, J. Mater. Chem. C **5** [41], 10701 (2017).
4) S. Ju, B. Liang, J.-Z. Wang, Y. Shi and S.-L. Li, Opt Commun **428**, 258 (2018).
5) T. Mahmoudi, Y. Wang and Y.-B. Hahn, Nano Energy **47**, 51 (2018).
6) S. K. Behura, C. Wang, Y. Wen and V. Berry, Nat Photonics **13** [5], 312 (2019).
7) X. Kong, L. Zhang, B. Liu, H. Gao, Y. Zhang, H. Yan and X. Song, RSC Adv **9** [2], 863 (2019).
8) Y. Mo, X. Deng, P. Liu, J. Guo, W. Wang and G. Li, Materials Science and Engineering: R: Reports **152**, 100711 (2023).
9) X. Li, H. Zhu, K. Wang, A. Cao, J. Wei, C. Li, Y. Jia, Z. Li, X. Li and D. Wu, Advanced Materials **22** [25], 2743 (2010).
10) X. Miao, S. Tongay, M. K. Petterson, K. Berke, A. G. Rinzler, B. R. Appleton and A. F. Hebard, Nano Lett **12** [6], 2745 (2012).
11) E. Shi, H. Li, L. Yang, L. Zhang, Z. Li, P. Li, Y. Shang, S. Wu, X. Li, J. Wei, K. Wang, H. Zhu, D. Wu, Y. Fang and A. Cao, Nano Lett **13** [4], 1776 (2013).
12) T. Cui, R. Lv, Z. H. Huang, S. Chen, Z. Zhang, X. Gan, Y. Jia, X. Li, K. Wang, D. Wu and F. Kang, J Mater Chem A Mater **1** [18], 5736 (2013).
13) Y. Song, X. Li, C. Mackin, X. Zhang, W. Fang, T. Palacios, H. Zhu and J. Kong, Nano Lett **15** [3], 2104 (2015).
14) L. Lancellotti, E. Bobeico, A. Capasso, E. Lago, P. Delli Veneri, E. Leoni, F. Buonocore and N. Lisi, Solar Energy **127**, 198 (2016).
15) A. Alnuaimi, I. Almansouri, I. Saadat and A. Nayfeh, Solar Energy **164**, 174 (2018).
16) L. Lancellotti, E. Bobeico, A. Castaldo, P. Delli Veneri, E. Lago and N. Lisi, Thin Solid Films **646**, 21 (2018).
17) M. A. Rehman, I. Akhtar, W. Choi, K. Akbar, A. Farooq, S. Hussain, M. A. Shehzad, S.-H. Chun, J. Jung and Y. Seo, Carbon **132**, 157 (2018).
18) M. Kim, M. A. Rehman, K. M. Kang, Y. Wang, S. Park, H. S. Lee, S. B. Roy, S. H. Chun, C. A. Singh, S. C. Jun and H. H. Park, Appl Mater Today **26**, 101267 (2022).
19) H. Al Busaidi, A. Suhail, D. Jenkins and G. Pan, Carbon Trends **10**, 100247 (2023).
20) E. Von Hauff, Journal of Physical Chemistry C **123** [18], 11329 (2019).
21) J. Bisquert and M. Janssen, J Phys Chem Lett **12** [33], 7964 (2021).
22) M. Zhong, D. Xu, X. Yu, K. Huang, X. Liu, Y. Qu, Y. Xu and D. Yang, Nano Energy **28**, 12 (2016).

23) J. Zhao, F.-J. Ma, K. Ding, H. Zhang, J. Jie, A. Ho-Baillie and S. P. Bremner, Appl Surf Sci **434**, 102 (2018).

24) M. A. Rehman, S. B. Roy, I. Akhtar, M. F. Bhopal, W. Choi, G. Nazir, M. F. Khan, S. Kumar, J. Eom, S.-H. Chun and Y. Seo, Carbon N Y **148**, 187 (2019).

25) C. Yim, N. McEvoy and G. S. Duesberg, Appl Phys Lett **103** [19], 193106 (2013).

26) S. Riazimehr, M. Belete, S. Kataria, O. Engström and M. C. Lemme, Adv Opt Mater **8** [13], 2000169 (2020).

27) I. Matacena, P. Guerriero, L. Lancellotti, E. Bobeico, N. Lisi, R. Chierchia, P. Delli Veneri and S. Daliento, Solar Energy **226**, 1 (2021).

28) Y. Ono and H. Im, Jpn J Appl Phys (2023) [ DOI:10.35848/1347-4065/acca57].

29) I. Mora-Seró, G. Garcia-Belmonte, P. P. Boix, M. A. Vázquez and J. Bisquert, Energy Environ Sci **2** [6], 678 (2009).

30) D. A. Neamen, *Semiconductor physics and devices : basic principles* (McGraw-Hill, 2012).

31) Wendy. Middleton, *Reference Data for Engineers, 9th Edition* (2001).

32) Bondarenko A S and Ragoisha G A, *Progress in Chemometrics Research* (Nova Science Publishers, New York, 2005).

Supplementary data

# Capacitance Characterization of Graphene/n-Si Schottky Junction Solar Cell with MOS Capacitor

Masahiro Teraoka, Yuzuki Ono, Hojun Im[*]

*Graduate School of Science and Technology, Hirosaki University, Hirosaki 036-8561, Japan*

[*]Corresponding author. Email: hojun@hirosaki-u.ac.jp (H. J. Im)

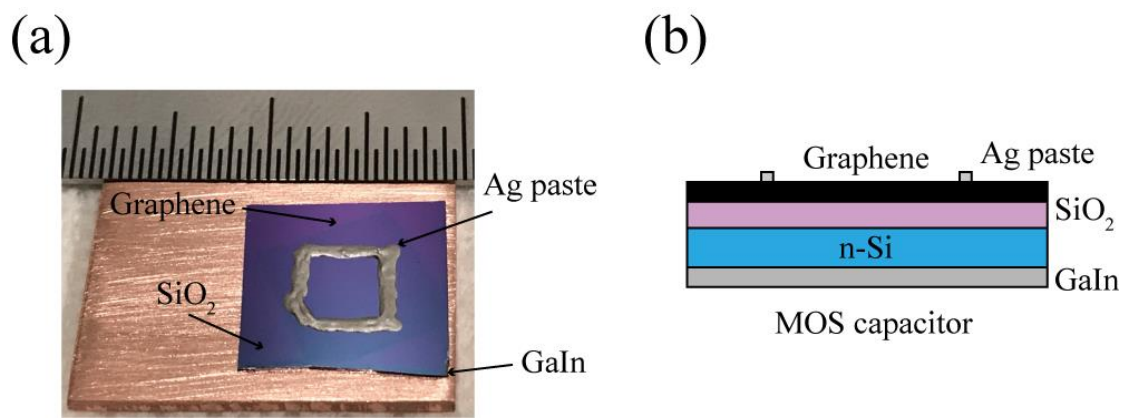

Fig. S1. (a) Photograph of the MOS capacitor, which has the same geometry as the GSSC except for the absence of the active area. (b) Schematic of the MOS capacitor in a side view.

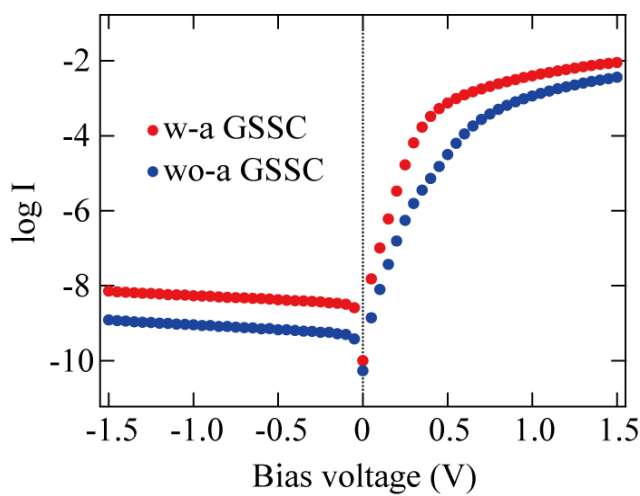

Fig. S2. I-V characteristics of the w-a and wo-a GSSCs in the semi-log scale.

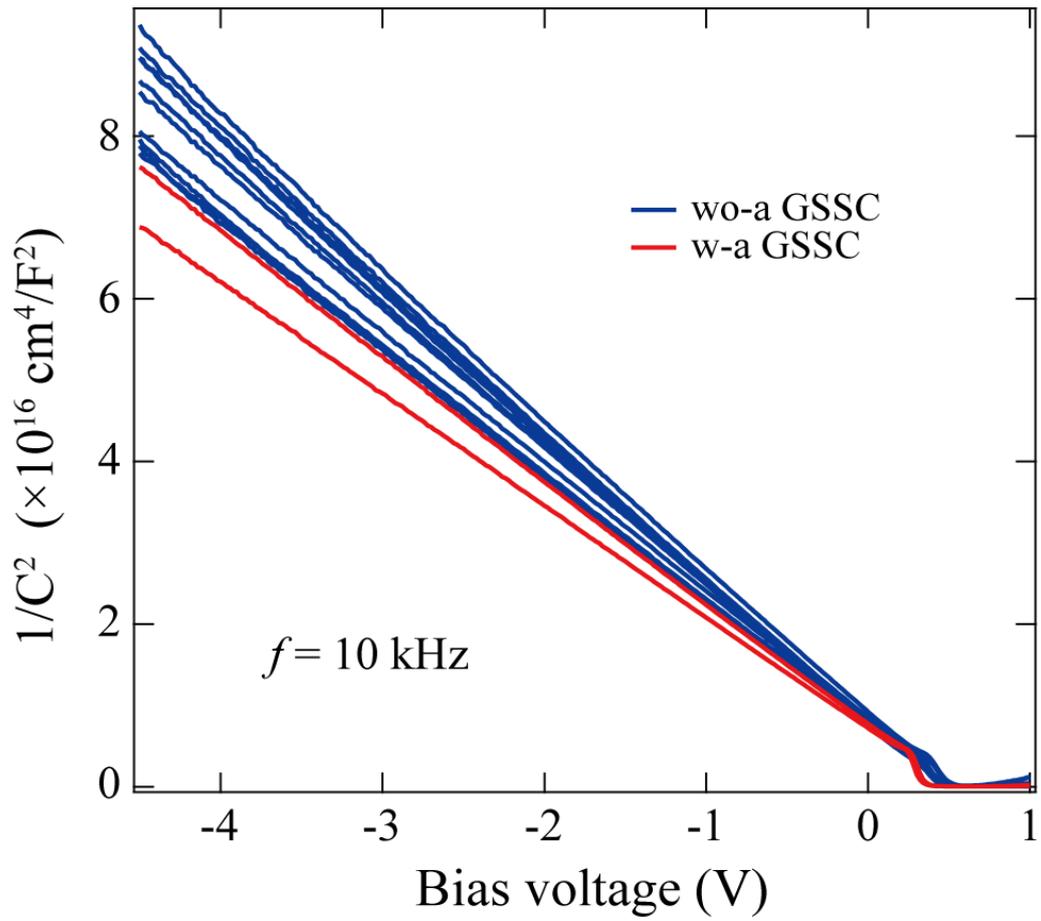

Fig. S3. Mott-Schottky plots obtained from $C_{Sch}$s for 12 GSSCs, including the data of Fig. 3(b).

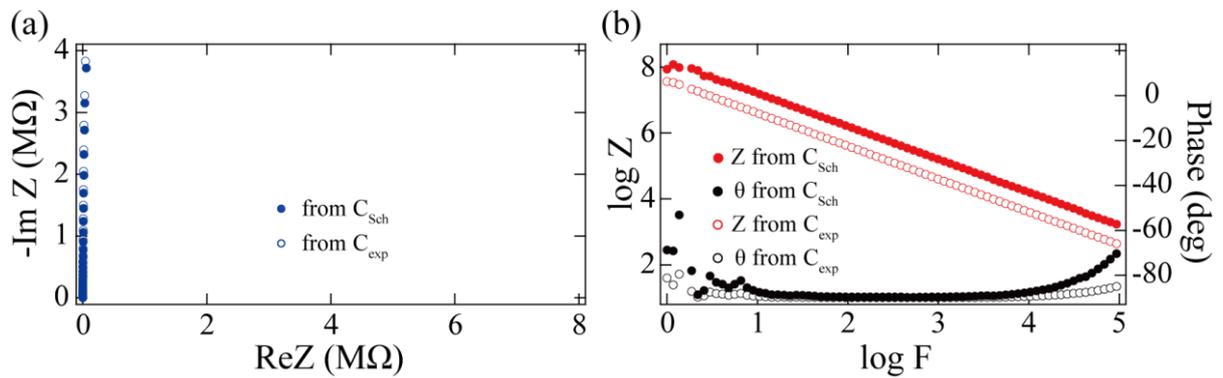

Fig. S4. (a) Nyquist and (b) Bode plots of the w-a and wo-a GSSCs obtained from the $C_{exp}$ and $C_{Sch}$.

16